# Gate-Modulated Quantum Interference Oscillations in Sb-Doped $Bi_2Se_3$ Topological Insulator Nanoribbon

Tae-Ha Hwang[1], Hong-Seok Kim[1], Yasen Hou[2], Dong Yu[2], Yong-Joo Doh[1,*]

[1]Department of Physics and Photon Science, Gwangju Institute of Science and Technology (GIST), Gwangju 61005, Korea

[2]Department of Physics, University of California, Davis, CA 95616, USA

**Corresponding Authors**

[*]E-mail: yjdoh@gist.ac.kr

# ABSTRACT

Topological insulator nanoribbons (TI NRs) provide a useful platform to explore the phase-coherent quantum electronic transport of topological surface states, which is crucial for the development of topological quantum devices. When applied with an axial magnetic field, the TI NR exhibits magnetoconductance (MC) oscillations with a flux period of $h/e$, *i.e.*, Aharonov–Bohm (AB) oscillations, and $h/2e$, *i.e.*, Altshuler-Aronov-Spivak (AAS) oscillations. Herein, we present an extensive study of the AB and AAS oscillations in Sb-doped $Bi_2Se_3$ TI NR as a function of the gate voltage, revealing phase-alternating topological AB oscillations. Moreover, the ensemble-averaged fast Fourier transform analysis on the $V_g$-dependent MC curves indicates the suppression of the quantum interference oscillation amplitudes near the Dirac point, which is attributed to the suppression of the phase coherence length within the low carrier density region. The weak antilocalization analysis on the perpendicular MC curves confirms the idea of the suppressed coherence length near the Dirac point in the TI NR.

## I. INTRODUCTION

Topological insulators (TIs) are bulk insulators including spin-resolved gapless surface states, which are topologically protected through time-reversal symmetry [1,2]. Because the spin orientation of the surface electrons is locked perpendicular to their translational momentum owing to a strong spin-orbit interaction in the TIs, the topological surface state has a spin-helical nature. In the form of a nanoribbon (NR) or nanowire structure, the quantum confinement effect along the circumference of the TI NR modifies the topological surface band to discrete one-dimensional (1D) subbands with a 1D band gap, $\Delta_{1D} = hv_F/L_p$, where $h$ is the Planck constant, $v_F$ is the Fermi velocity, and $L_p$ is the perimeter length of the TI NR [3,4]. As a combination of a nontrivial topological order and coherent electronic transport, the TI NR provides a useful platform for exploring topological quantum computing [5-7] and spintronics [8,9].

When we apply an external magnetic field parallel to the NR axis, the dispersion relation for the 1D subbands in the TI NR is as follows [4,10]:

$$E(n,k,\Phi) = \pm hv_F \sqrt{\left(\frac{k}{2\pi}\right)^2 + \left(\frac{n+1/2-\Phi/\Phi_0}{L_p}\right)^2}, \qquad (1)$$

where $k$ is the wave number along the axial direction, $n$ is an integer, $\Phi$ is the magnetic flux threading the cross-sectional area of the NR, $\Phi_0 = h/e$ is the flux quantum, and $e$ is the elementary charge. Here, the half-integer term of 1/2 is due to the spin Berry phase of $\pi$ owing to the spin-momentum locking in the topological surface states, and $\Phi/\Phi_0$ is due to the Aharonov–Bohm (AB) phase [11] added to the electronic wave function along the circumference of the TI NR. It is expected from the flux-dependent dispersion relation that the density of states of the 1D subbands will be modulated periodically with the axial magnetic flux, resulting in magnetoconductance (MC) oscillations with a period of $\Phi_0 = h/e$ [4], which are known as AB oscillations [12]. It is also expected that the conductance maxima owing to the AB oscillations in the TI NR can occur at integer (0-AB) or half-integer flux quanta ($\pi$-AB oscillations), depending on the location of the Fermi level, which becomes a characteristic feature of topological AB oscillations [3].

Topological AB oscillations have been observed experimentally in various TI NRs of $Bi_2Se_3$ [4,7,12,13], $Bi_2Te_3$ [10,14,15], SnTe [16], $Sb_2Te_3$ [17], and Dirac semimetal NRs [18,19]. There have been many reports [4,12,14,15,18,20] on the observations of $h/e$-periodic AB oscillations concurrently with the $h/2e$-periodic oscillations, which are attributed to the weak antilocalization (WAL) effect along the circumference of TI NR in the diffusive regime [12], which is also called Altshuler–Aronov–Spivak (AAS) oscillations [21]. The observations of quantum oscillations in the TI NR suggest that TI NRs are promising mesoscopic systems for studying quantum interference effects combined with topological surface states. Few studies [10,20], however, have provided systematic evidence of the 0 and $\pi$ phase-alternating behaviors of AB oscillations in the TI NR as a function of the gate voltage $V_g$. Moreover, although the AB oscillation amplitudes were theoretically expected to be enhanced near the Dirac point [3,22], the suppression of both AB and AAS oscillations was observed at near the Dirac point without a clear explanation [14].

In this report, we present an extensive experimental study of AB and AAS oscillations obtained from Sb-doped $Bi_2Se_3$ TI NRs by varying the gate voltages. Under the application of an axial magnetic field, we clearly observed a $V_g$-dependent phase alternation of the $h/e$-periodic AB oscillations, superimposed with additional $h/2e$-periodic AAS oscillations. The $V_g$-sensitive axial MC curves were analyzed using the ensemble-averaged fast Fourier transform (FFT) method to avoid any confusion from sample-specific features, resulting in the suppression of the AB and AAS oscillation amplitudes at near the Dirac point. We suggest that these suppressions of quantum interference oscillations are due to a suppression of the phase coherence length, $L_\phi$, in the low carrier density region, thereby increasing the effective number of AB loops in the TI NR. An additional WAL analysis of the perpendicular MC data confirms the suppression of $L_\phi$ near the Dirac point. Our observations provide direct evidence of the topological AB oscillations in the TI NR and emphasize the importance of the phase coherence length in understanding the quantum electronic transport in the TI NR.

## II. EXPERIMENT

Sb-doped $Bi_2Se_3$ NRs were synthesized using a chemical vapor deposition method using Au nanoparticles as a catalyst in a horizontal tube furnace. More detailed information regarding NR growth can be found elsewhere [23]. Energy-dispersive X-ray spectroscopy of TI NRs revealed atomic percentages of approximately 36.0%, 5.5%, and 58.5% for Bi, Sb, and Se, respectively [20]. After the growth of TI NRs, individual NRs were transferred onto a highly electron-doped silicon substrate covered with a 300-nm thick $SiO_2$ layer. The silicon substrate was used as the back gate. Source and drain electrodes were formed using electron-beam lithography with a field-emission scanning electron microscope (Tescan Mira-3), followed by the deposition of a Ti(10 nm)/Au(200 nm) film using an electron-beam evaporator. Prior to the metal deposition, the surface of the TI NR was treated with oxygen plasma to remove the residual electron-beam resist, and then cleaned with a 6:1 buffered oxide etch solution for 7 s to remove the native oxide layer [7]. All electrical transport measurements of the TI NR devices were conducted using a conventional lock-in technique and a four-probe method applied in a closed-cycle $^4$He cryostat (Seongwoo Instruments Inc.), which has a base temperature of 2.4 K. Low-pass RC and $\pi$ filters were connected in-series with the measurements, thereby reducing environmental noise.

## III. RESULTS AND DISCUSSION

A scanning electron microscopy image of the Sb-doped $Bi_2Se_3$ NR devices is shown in Fig. 1a. The width and thickness of the TI NR were $w = 325$ nm and $t = 90$ nm, respectively, whereas the total length was approximately $L_{tot} = 150$ μm. Device 1 (D1) and Device 2 (D2) indicate two different NR segments between electrode pairs of 2-3 and 3-4, respectively. The channel lengths were $L_{ch} = 5$ and 70 μm for D1 and D2, respectively. The axial magnetic field $B_{axial}$ was applied along the NR axis, as shown in the figure. The square resistance, $R_s$, of D1 and D2 is depicted as a function of gate voltage $V_g$ at $T = 2.6$ K, as shown in Fig. 1b. It has been inferred that the Dirac point, which is the gate voltage corresponding to the maximum $R_s$, is located at $V_{DP} = -39$ (−25) V for D1 (D2). The difference in $V_{DP}$ of the two devices is attributed to an inhomogeneous doping profile or an impurity distribution in the TI NR.

Figure 1c shows representative MC curves of the TI NR under the application of an axial magnetic field at low temperatures. It is clearly shown that regular periodic MC oscillations are superposed on a negative parabolic background owing to a classical origin [24]. This feature was commonly observed for both D1 and D2 devices except their oscillation periods. After subtracting the smooth background, the conductance difference, $\Delta G$, was obtained as a function of $B_{axial}$ for both devices, as shown in Fig. 1d. Oscillation periods of $\Delta B_{axial} = 0.17$ T and 0.08 T were obtained for D1 and D2 from the figure, respectively. Using the geometrical parameters of the TI NR, the corresponding magnetic flux through a cross-sectional area of the TI NR turned out to be $\Delta\Phi = wt\Delta B_{axial} = 1.04$ $\Phi_0$ for D1 and 0.49 $\Phi_0$ for D2, considering a 5-nm thick native oxide layer residing on the surface of the TI NR [15,25].

The FFT analysis of the $\Delta G$ curve reveals the coexistence of two oscillations for both devices D1 and D2 (see Fig. 1e). The first FFT peak at the axial field frequency of $f_B = 6.0$ T$^{-1}$ corresponds to the $h/e$-periodic oscillations, whereas the second peak at $f_B = 12.0$ T$^{-1}$ corresponds to the $h/2e$-periodic oscillations. From the FFT amplitudes, we noted that the MC oscillations in D1 are dominated by $h/e$-periodic AB oscillations, whereas the MC oscillations in D2 are dominated by $h/2e$-periodic oscillations. The difference in the dominant quantum interference oscillations of the two devices formed on the same TI NR indicates that the quantum transport governing each device strongly depends on the channel length of the TI NR device [20]. More conclusive results, however, require information regarding the $V_g$ dependence of the $\Delta G$ curve because the MC curves are extremely sensitive to $V_g$.

Figure 2a shows color plots of $\Delta G$ obtained from D1 as a function of $V_g$ and $B_{axial}$. The $h/e$-periodic oscillations of $\Delta G$ are clearly observed at near $V_g = -16$ V, and $\Delta G(B_{axial} = 0)$ values alternate between the maxima and minima depending on $V_g$, exhibiting an elongated checkerboard pattern. When the MC maxima occur at integer (half-integer) multiples of $\Phi_0$, they are called 0-phase ($\pi$-phase) AB oscillations [10]. The $V_g$-dependent phase alternating behavior of AB oscillations is attributed to the formation of 1D sub-bands of the topological surface states in the TI NR [3,4], which becomes a characteristic feature of the topological AB oscillations. We note that $h/2e$-periodic oscillations occur within the $V_g$ region, where two AB oscillations with different phases overlap. When the Fermi energy approaches the Dirac point, $\Delta G$ decreases and the checkerboard pattern diminishes.

Because the phase and amplitude of the individual MC curves are sensitive to $V_g$, we conducted an ensemble-averaged FFT analysis on the $\Delta G(B_{axial})$ curves to examine the gate voltage dependence of the quantum interference oscillation amplitudes. The FFT spectrum in

Fig. 2b was obtained by averaging 41 traces of the FFT spectra, wherein each spectrum was computed from $\Delta G$ curves measured at different values of $V_g$ with a constant increment of $\Delta V_g$ = 50 mV and the average gate voltage, $<V_g>$. The peak heights for both $h/e$- and $h/2e$-periodic oscillations are displayed as a function of $<V_g>$ in Fig. 2c. It should be noted that the FFT amplitudes of both quantum interference oscillations are comparable in our experimental range of $V_g$, indicating that D1 is within the crossover regime between the ballistic and diffusive quantum transport in the TI NR [20]. Moreover, the FFT amplitudes are maximized at $V_g = -16$ V, while decreasing at near the Dirac point and at $V_g = 0$ V. The physical mechanism underlying the $V_g$-dependent FFT amplitudes will be discussed later.

The color plots of $\Delta G(B_{axial}, V_g)$ obtained from D2 are shown in Fig. 2d. At near $V_g$ = −16 V, $h/2e$-periodic oscillations dominate the MC curves, especially within the low $B_{axial}$ regime, whereas $h/e$-periodic oscillations recover at higher fields of $|B_{axial}| > 0.5$ T. A similar behavior has been observed in other TI NRs [4] and core/shell semiconductor nanowires [26], which has been attributed to a suppression of coherent AAS oscillations owing to the broken time-reversal symmetry induced by the higher $B_{axial}$ field. The $h/e$-periodic AB oscillations, however, are due to a coherent forward scattering and thus persist at up to higher $B_{axial}$ fields. At near the Dirac point, $\Delta G$ decreases, although the $h/2e$-periodic oscillations are still visible. After conducting an ensemble-averaged FFT analysis, we obtained $V_g$-dependent FFT amplitudes and the peak heights of the $h/e$- and $h/2e$-periodic oscillations are shown in Figs. 2e and 2f, respectively. It should also be noted that the peak height was maximized at near $V_g$ = −6 and −11 V for $h/e$- and $h/2e$-periodic oscillations, respectively, while decreasing at near the Dirac point and at above $V_g = 0$ V. In contrast to D1, the maximum peak height of the $h/2e$-periodic oscillations in D2 was almost three-times larger than that of the $h/e$-periodic oscillations, indicating that D2 was within the diffusive quantum transport regime.

TI NR, analogous to a hollow metallic cylinder, can be considered as a series array of AB loops [27] and the number of AB loops is given by $N_{AB} = L_{ch}/L_\phi$, where $L_\phi$ is the phase coherence length [28]. It is well known that $L_\phi$ can be obtained from the perpendicular MC data of the TI NR [15,19]. When a magnetic field $B$ is applied perpendicular to the top surface of the TI NR, negative MC behavior is expected from the WAL effect owing to a destructive quantum interference between a pair of time-reversed electron paths formed on the top and bottom surfaces of the TI NR. Figure 3a shows the measurement configuration for the perpendicular MC data (see the inset) and $R_s$ versus $V_g$ curve of device D3 with $L_{ch} = 5$ μm formed on a TI NR with $w = 607$ nm and $t = 90$ nm. Under application of a perpendicular $B$ field, D3 exhibited a negative magnetoconductivity, $\sigma(B) = 1/R_s(B)$, and its differences from the zero-field value, $\delta\sigma(B) = \sigma(B) - \sigma(0)$, are shown in Fig. 3b as a color plot depending on the $V_g$ and $B$ fields.

To reduce the universal conductance fluctuations [29,30] in the TI NR, $\delta\sigma(B)$ was averaged over 11 traces obtained at different values of $V_g$ with a constant increment of $\Delta V_g = 1$ V, as shown in Fig. 3c (symbols). The resulting $\delta\sigma(B)$ curves were fitted to the two-dimensional WAL theory [31], *i.e.*, the so-called Hikami–Larkin–Nagaoka (HLN) model, which is

$$\delta\sigma(B) = \frac{-\alpha e^2}{2\pi^2 \hbar}\left[\ln\left(\frac{\hbar}{4eBL_\varphi^2}\right) - \Psi\left(\frac{1}{2} + \frac{\hbar}{4eBL_\varphi^2}\right)\right], \qquad (2)$$

where $\alpha$ is a prefactor, $\hbar$ is the reduced Planck's constant, and $\Psi$ is the digamma function. The fitting results, shown by the lines in Fig. 3c, agree well with the experimental data, and the fitting parameters, $\alpha$ and $L_\phi$, are depicted as a function of $V_g$ in Fig. 3d. Note that the coherence length is maximized as $L_\phi = 560$ nm at $V_g = 10$ V and then decreases at near the Dirac point and at above $V_g = 10$ V. The suppression of $L_\phi$ within the low carrier density regime in TI has been observed in previous reports [32,33], which is attributed to an enhanced electron–electron interaction owing to the reduced screening effect [32]. In addition, the prefactor is relatively insensitive to $V_g$ and is averaged as approximately $\alpha = -1.1$, implying that two topological surface states exist in the top and bottom surfaces of the TI NR [34].

We now return to our observations of $V_g$-dependent quantum interference oscillations and discuss their underlying physics. It is well known that a series array of metallic AB loops suppresses the AB oscillation amplitude, which is proportional to $N_{AB}^{-3/2} = (L_{ch}/L_\phi)^{-3/2}$, owing to the random phase shift of the AB oscillations in each loop [28]. Here, the effect of the connecting leads between the AB loops, $N_{AB}^{-1}$, was considered along with the stochastic effect of the loops, $N_{AB}^{-1/2}$. As $L_\phi$ decreases near the Dirac point in the TI NR, as already shown in Fig. 3d, it is expected that the AB amplitude will be suppressed in the low-density regime and/or for an NR segment with a longer $L_{ch}$, which is consistent with our observations in Figs. 2c and 2f. When we compare the maximum and minimum values of $L_\phi$ in Fig. 3d, it is estimated that $N_{AB}$ at near the Dirac point increases by ~ 1.9 times, as compared with the $N_{AB}$ value at the maximum $L_\phi$. As a result, the $h/e$-periodic oscillation amplitude at near the Dirac point will be suppressed to approximately 39% of its maximum value, which is close to our experimental values of 43% and 45% obtained from D1 and D2, respectively.

The $h/2e$-periodic AAS oscillations, however, yield a zero phase shift owing to the phase coherence between a pair of time-reversed paths formed by sequential disorder scattering. Instead, the AAS oscillation amplitude is proportional to $\exp(-2L_p/L_\phi)$, where $L_p$ is the perimeter length of the TI NR [25,28]. Because $L_\phi$ at the Dirac point becomes almost half of the maximum value, the suppression of the AAS oscillation amplitudes is expected at near the Dirac point. Additional decreasing behaviors of the $h/e$- and $h/2e$-periodic oscillation amplitudes at above the $V_g$ values corresponding to their amplitude peaks are attributed to the contribution of the bulk conduction band, which enhances the inelastic electron–electron scattering, resulting in a reduction of $L_\phi$. This inclusion of the bulk property will be plausible when the Fermi level is located inside the bulk conduction band within the high $V_g$ region.

## IV. CONCLUSION

In summary, we extensively studied the $V_g$ dependence of the $h/e$- and $h/2e$-periodic MC oscillations in Sb-doped $Bi_2Se_3$ TI NRs. Alternating behaviors of the 0- and π-phase AB oscillations were observed as a function of $V_g$, as expected from the theory of topological AB oscillations in the TI NR. The ensemble-averaged FFT analysis of the MC curves revealed that an amplitude suppression at near the Dirac point was commonly observed for both the AB and AAS oscillations, which was attributed to the suppression of the phase coherence length within the low carrier concentration region. The WAL analysis of the perpendicular MC data confirmed the $V_g$ dependence of the phase coherence length in the TI NR.

## ACKNOWLEDGMENTS

This research was supported by the NRF of Korea through the Basic Science Research Program (2018R1A3B1052827) and the U.S. National Science Foundation (Grant DMR-1838532).

# REFERENCES

[1] L. Fu, C. L. Kane, and E. J. Mele, Phys. Rev. Lett. **98**, 106803 (2007).

[2] M. Z. Hasan and C. L. Kane, Rev. Mod. Phys. **82**, 3045 (2010).

[3] J. H. Bardarson, P. W. Brouwer, and J. E. Moore, Phys. Rev. Lett. **105**, 156803 (2010).

[4] S. S. Hong, Y. Zhang, J. J. Cha, X.-L. Qi, and Y. Cui, Nano Lett. **14**, 2815 (2014).

[5] J. Manousakis, A. Altland, D. Bagrets, R. Egger, and Y. Ando, Phys. Rev. B **95**, 165424 (2017).

[6] M. Kim *et al.*, Nat. Commun. **10**, 4522 (2019).

[7] N.-H. Kim, H.-S. Kim, Y. Hou, D. Yu, and Y.-J. Doh, Curr. Appl. Phys. **20**, 680 (2020).

[8] C. H. Li *et al.*, Nat. Nanotechnol. **9**, 218 (2014).

[9] T.-H. Hwang, H.-S. Kim, H. Kim, J. S. Kim, and Y.-J. Doh, Curr. Appl. Phys. **19**, 917 (2019).

[10] L. A. Jauregui, M. T. Pettes, L. P. Rokhinson, L. Shi, and Y. P. Chen, Nat. Nanotechnol. **11**, 345 (2016).

[11] Y. Aharonov and D. Bohm, Phys. Rev. **115**, 485 (1959).

[12] H. Peng *et al.*, Nat. Mater. **9**, 225 (2010).

[13] S. Cho *et al.*, Nat. Commun. **6**, 7634 (2015).

[14] F. Xiu *et al.*, Nat. Nanotechnol. **6**, 216 (2011).

[15] H.-S. Kim *et al.*, Curr. Appl. Phys. **16**, 51 (2016).

[16] M. Safdar *et al.*, Nano Lett. **13**, 5344 (2013).

[17] Y. C. Arango *et al.*, Sci. Rep. **6**, 29493 (2016).

[18] L.-X. Wang, C.-Z. Li, D.-P. Yu, and Z.-M. Liao, Nat. Commun. **7**, 10769 (2016).

[19] J. Kim *et al.*, ACS Nano **10**, 3936 (2016).

[20] H.-S. Kim *et al.*, arXiv:2008.08421 (2020).

[21] B. L. Al'Tshuler, A. G. Aronov, and B. Z. Spivak, J. Exp. Theor. Phys. **33**, 94 (1981).

[22]   J. Dufouleur, E. Xypakis, B. Büchner, R. Giraud, and J. H. Bardarson, Phys. Rev. B **97**, 075401 (2018).

[23]   Y. Hou *et al.*, Nat. Commun. **10**, 5723 (2019).

[24]   J.-W. Chang *et al.*, Appl. Phys. Express **6**, 085201 (2013).

[25]   J. Dufouleur *et al.*, Physical Review Letters **110**, 186806 (2013).

[26]   M. Jung *et al.*, Nano Lett. **8**, 3189 (2008).

[27]   C. P. Umbach, C. Van Haesendonck, R. B. Laibowitz, S. Washburn, and R. A. Webb, Phys. Rev. Lett. **56**, 386 (1986).

[28]   S. Washburn and R. A. Webb, Adv. Phys. **35**, 375 (1986).

[29]   P. A. Lee and A. D. Stone, Phys. Rev. Lett. **55**, 1622 (1985).

[30]   Y.-J. Doh, A. L. Roest, E. P. A. M. Bakkers, S. De Franceschi, and L. P. Kouwenhoven, J. Korean Phy. Soc. **54**, 135 (2009).

[31]   S. Hikami, A. I. Larkin, and Y. Nagaoka, Prog. Theor. Exp. Phys. **63**, 707 (1980).

[32]   J. Chen *et al.*, Phys. Rev. Lett. **105**, 176602 (2010).

[33]   J. Lee, J. Park, J.-H. Lee, J. S. Kim, and H.-J. Lee, Phys. Rev. B **86**, 245321 (2012).

[34]   J. J. Cha *et al.*, Nano Lett. **12**, 1107 (2012).

**FIGURE CAPTIONS**

**Fig. 1.** (a) Scanning electron microscopy image of Sb-doped $Bi_2Se_3$ NR devices. Each electrode is labeled with a number. The axial field, $B_{axial}$, was applied parallel to the NR axis. (b) Square resistance versus $V_g$ curves of D1 and D2 at $T = 2.6$ K. (c) MC curves obtained from D1 (red curve at left axis) and D2 (blue at right axis). (d) The conductance difference $\Delta G(B_{axial})$ curves obtained from D1 and D2, after subtracting the smooth background. (e) Normalized FFT spectra of D1 and D2. The two arrows indicate the peaks corresponding to the $h/e$- and $h/2e$-periodic oscillations, respectively.

**Fig. 2.** (a) Color plots of $\Delta G(B_{axial}, V_g)$ data of D1 for different $<V_g>$ at $T = 2.6$ K. (b) Ensemble-averaged FFT spectra for different $<V_g>$ of D1 (see text). (c) $V_g$ dependence of the FFT peak heights corresponding to the $h/e$- (circles) and $h/2e$-periodic oscillations (square symbols) of D1. (d) Color plots of $\Delta G(B_{axial}, V_g)$ data of D2 for different $<V_g>$. (e) Ensemble-averaged FFT spectra for different $<V_g>$ of D2. (f) $V_g$ dependence of the FFT peak heights corresponding to the $h/e$-(circles) and $h/2e$-periodic oscillations (square symbols) of D2.

**Fig. 3.** (a) Square resistance versus $V_g$ curve of D3 at $T = 3.3$ K. Inset: Optical microscopy image of D3 and a schematic of four-point measurement method. A magnetic $B$ field was applied perpendicular to the substrate. Scale bar indicates 5 μm. (b) Color plot of $\delta\sigma(B, V_g)$ data. (c) Ensemble-averaged $\delta\sigma(B)$ data (symbols) and HLN fitting lines with different $<V_g>$ (see text). (d) $V_g$ dependence of the HLN fitting parameters of the phase coherence length ($L_\phi$ at left axis) and a prefactor $\alpha$ (right axis).

# Figure 1

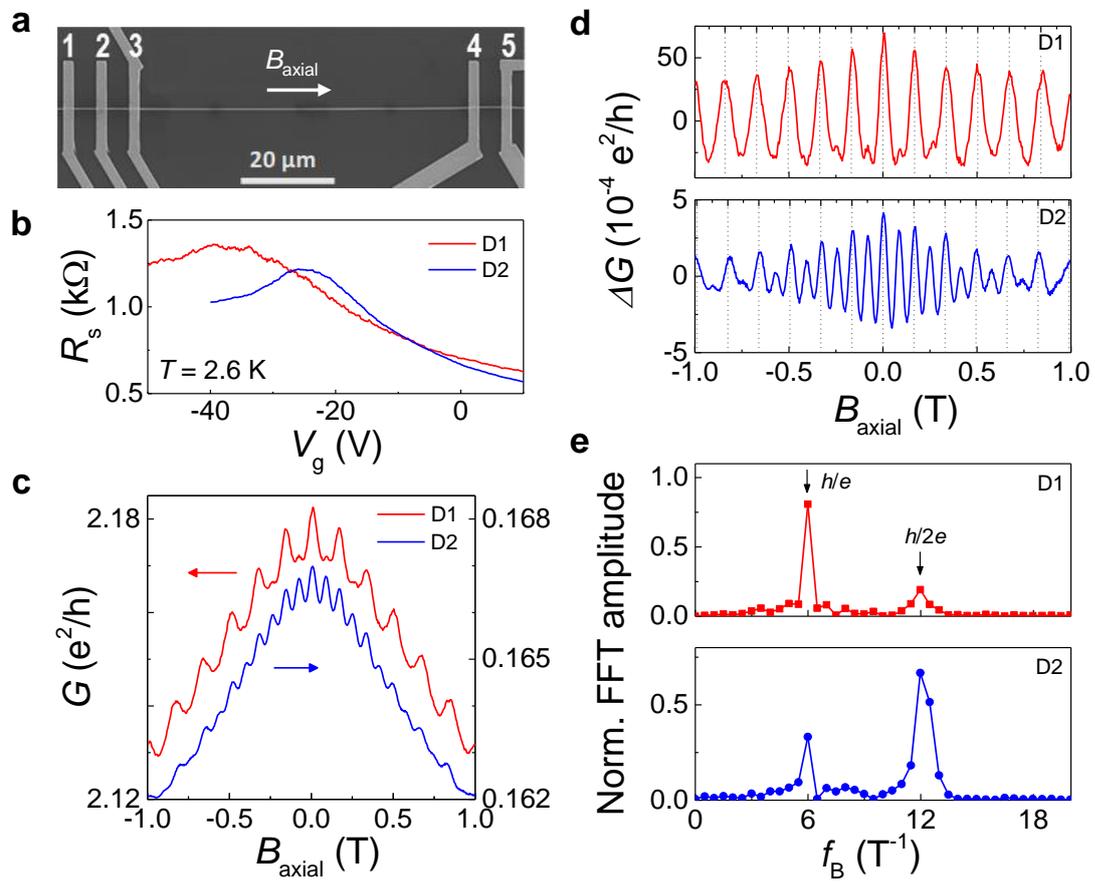

# Figure 2

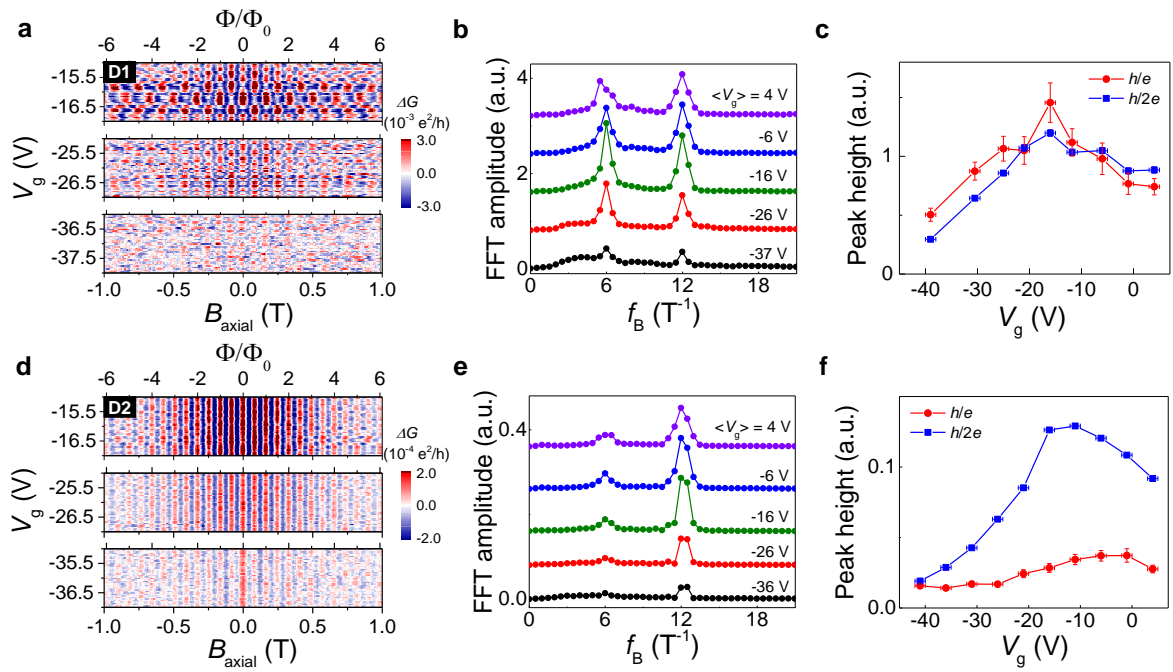

**Figure 3**

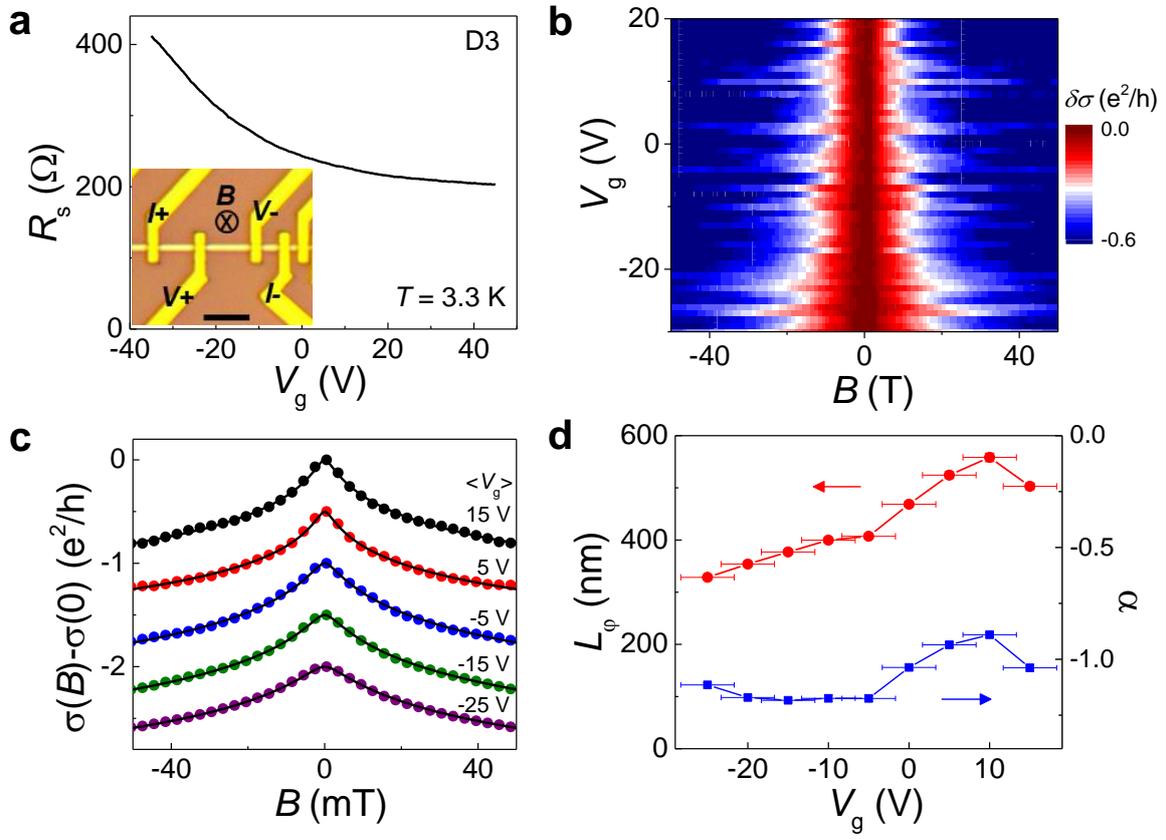